\documentclass[prb,twocolumn,superscriptaddress]{revtex4-1}
\usepackage{amsmath,amssymb,graphicx,color,bm,epstopdf,float}

\begin{document}

\title{Electronic and Magnetic Properties of Graphite Quantum Dots}
\author{Hazem Abdelsalam }

\affiliation{University of Picardie, Laboratory of Condensed Matter
Physics, Amiens, 80039, France} \affiliation{Department of
Theoretical Physics, National Research Center, Cairo,12622, Egypt}

\author{T. Espinosa-Ortega}
\affiliation{Division of Physics and Applied Physics, Nanyang
Technological University 637371, Singapore}
\author{Igor Lukyanchuk}
\affiliation{University of Picardie, Laboratory of Condensed Matter
Physics, Amiens, 80039, France}

\begin{abstract}
We study the electronic and magnetic properties of multilayer
quantum dots (MQDs) of graphite in the nearest-neighbor
approximation of tight-binding model. We calculate the electronic
density of states and orbital susceptibility of the system as
function of the Fermi level location. We demonstrate that properties
of MQD depend strongly on the shape of the system, on the parity of
the layer number and on the form of the cluster edge. The special
emphasis is given to reveal the new properties with respect to the
monolayer graphene quantum dots (GQD). The most interesting results
are obtained for the triangular MQD with zig-zag edge at near-zero
energies. The asymmetrically smeared multi-peak feature is observed
at Dirac point within the size-quantized energy gap region, where
monolayer graphene flakes demonstrate the highly-degenerate
zero-energy state. This feature, provided by the edge-localized
electronic states results in the splash-wavelet behavior in
diamagnetic orbital susceptibility as function of energy.
\end{abstract}

\maketitle


\section{Introduction}

Rise of graphene certainly revived the interest to the classical
graphite systems,  presenting a wealth of not yet well understood
electronic and magnetic properties. The challenge is related to the
complicate semi-metallic multi-branch energy spectrum in the
vicinity of the half-field Fermi-level, caused by splitting of the
Dirac-cone graphene spectrum by the graphite-forming
inter-carbon-layer coupling. The point of special interest is the
crossover from graphene to graphite through the multilayer
structures with few number of layers. It is well known that such
systems may exhibit metallic or semiconductor behavior as a function
of the
number of layers and the stacking process \cite%
{Latil2006,Partoens2007,Santos2007, Finkenstadt2008,koshino2013},
characteristic that makes them highly appealing for gated controlled
electronic devices.

The magnetic properties of few-layer structure are even more intriguing\cite%
{Safran,Saito,Koshino1,Hirasawa,Castro}. The orbital magnetism of
the
odd-layer structures is reminiscent to that for the monolayer graphene\cite%
{Koshino1}. In particularly, a characteristic for graphene diamagnetic $%
\delta $-function singularity of susceptibility
\cite{McClure,Slonczewski 1958} appears at the Dirac point at $E=0$.
For even number of layers the
magnetic properties are more similar to those for bilayer graphene \cite%
{Koshino1} and the diamagnetic response has the weaker logarithmic
divergency. Such odd/even layer decomposition can give the
coexistence of
Dirac and normal carriers, observed in the pure graphite \cite%
{Lukyanchuk2004,Lukyanchuk2006,Lukyanchuk2009}.

Alongside, flake-like graphene quantum dots (GQD) have captured the
substantial attention of nanotechnology due to their unique optical
and
magnetic properties \cite%
{Ezawa,Zhang2008,Rossier2007,Wang2008,Wang2009,Potasz,Kosimov,Espinosa,Zarenia,Heiskanen,Espinosa2,Liu,Ominato,Ominato2013}%
. The new element here is the finite-size electron confinement,
resulted in  opening of the energy gap for the bulk delocalized
electronic states in
the vicinity of Dirac point \cite%
{Ezawa,Zhang2008,Rossier2007,Potasz,Kosimov,Heiskanen}. This gap,
however can be filled by the energy level of novel electronic
states, localized in the vicinity of the sample boundary
\cite{Ezawa,Zhang2008,Rossier2007}. For nanoscopic and even for
mesoscopic clusters these \textit{edge states} can play the decisive
role in electronic and magnetic properties of the flakes
\cite{Rossier2007,Wang2008,Wang2009,Ominato,Ominato2013}. The
situation can drastically depend on size and shape of the clusters.
Even the geometrical
structure of the edges (armchair vs zigzag) plays the important role \cite%
{Ezawa,Zhang2008,Heiskanen}. In general two types of edge states,
located nearby the Dirac point can be discerned, the zero energy
states (ZES) that are degenerate and located \textit{exactly} at
$E=0$ and the dispersed edge states (DES) that fill the low-energy
energy-spectra domain within the gap
and are symmetrically distributed with respect to $E=0$ \cite%
{Espinosa,Zarenia,Heiskanen,Espinosa2}

In this paper, we consider the electronic and magnetic properties of
finite-size multilayer quantum dots (MQD) that should generalize the
principal features of GQD in that sense as the described above
extended multilayer systems grasp the properties of graphene. In
particular we show that the edge state located close to $E=0$ are
again responsible for the principal electronic properties, but the
level arrangement inside the gap is more diverse as in GQD. To
stress the most prominent aspects we consider the characteristics
examples of MQD of hexagonal and of triangular shape having zigzag
edges. For calculations we use the approach of Tight Binding (TB)
model in the nearest neighbor (NN) approximation. For magnetic
properties we are mostly concentrated on the orbital diamagnetic
effects. The role of the spin-paramagnetic properties is briefly
discussed in the conclusion and will be studied elsewhere.

\label{sec:intro}

\section{The model}

\label{sec:Model}

Graphene is formed by a two-dimensional honeycomb lattice of carbon
atoms in which the conducting $\pi $-band electrons can be described
within the TB model as
\begin{equation}
H=\sum_{i}\varepsilon _{i}c_{i}^{\dag }c_{i}+\sum_{\left\langle
ij\right\rangle }(t_{ij}c_{i}^{\dag }c_{j}+{t}_{ji}{c}_{j}^{\dag
}{c}_{i}), \label{Htb_s}
\end{equation}%
where $c_{i}^{\dag }$ and $c_{i}$ are the creation and annihilation
electron
operators, $t_{ij}$ are the inter-site electron hopping elements and $%
\varepsilon _{i}$ is the on-site electron energy. The Hamiltonian (\ref%
{Htb_s}) can be extended for the multilayer systems by taking into
account the NN\ hopping between the adjacent layers. Five interlayer
coupling parameters $\gamma _{1}\ldots \gamma _{5}$ were introduced
by Slonczewski, Weiss and McClure \cite{Slonczewski 1958,McClure} as
the hopping parameters
for the graphite structure (Fig. 1a). Parameter $\gamma _{1}\simeq 0.39~%
\mathrm{eV}$ represents the coupling between the vertically aligned
$B_{N}$
and $A_{N+1}$ atoms (subscript index means the number of plane, shift $%
N\rightarrow N+1$ permutes A and B), parameter $\gamma _{3}\simeq 0.315~%
\mathrm{eV}$ describes the coupling between the shifted $A_{N}$ and
$B_{N+1}$ atoms and parameter $\gamma _{4}\simeq 0.044~\mathrm{eV}$
corresponds to the
coupling between the $A_{N}$ and $A_{N+1}$ and between the $B_{N}$ and $%
B_{N+1}$ atoms. Another two parameters, $\gamma _{2}$ and $\gamma
_{5}$ represent the coupling between the next-nearest neighboring
(NNN) layers.
Parameter $\gamma _{5}\simeq 0.04~\mathrm{eV}$ connects atoms $B_{N}$ and $%
B_{N+2}$ belonging to the same vertical line as atoms connected by
parameter $\gamma _{1}$ whereas parameter $\gamma _{2}\simeq
-0.02~\mathrm{eV}$ corresponds to another vertically aligned atoms
$A_{N}$ and $A_{N+2}$, with no intermediate atom between them.\ In
addition, the on-site electron energies $\epsilon_i$ of A and B
atoms in layers of MQD\ become different and described by the gap
parameters $\Delta =\pm 0.047~\mathrm{eV}$ \cite{koshino2013}
alternately.

\begin{figure}[h]
\includegraphics[width=.5\textwidth]{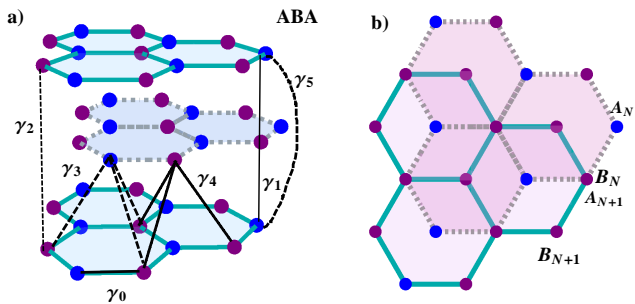}
\caption{ Coupling parameters for multilayer ABA carbon stacking
(a). Top view of the multilayer structure (b). The carbon atoms
belonging to sublattices $A_{N}$ and $B_{N}$ are shown by blue and
red colors.} \label{Fig1}
\end{figure}

\

We use the TB Hamiltonian (\ref{Htb_s}) to study the MQD\ of
triangular and hexagonal shape with zig-zag termination. We assume
that graphene layers are arranged in the graphite-type ABA stacking
as shown in Fig.\ref{Fig1}. The electronic energy levels of MQD,
$E_{n}$\ and corresponding DOS are found from the TB Hamiltonian
(\ref{Htb_s}) with interlayer hopping. In current article we
consider mostly the NN\ layer coupling, neglecting the effects of
$\gamma _{2}$, $\gamma _{5}$ and $\Delta $.\ The effect of a
$c$-directed magnetic field is accounted by using the Peierls
substitution for the hopping matrix elements $t_{ij}$ between atomic
sites $\mathbf{r}_{i}$ and $\mathbf{r}_{j}$
\begin{equation}
t_{ij}\rightarrow t_{ij}^{P}=t_{ij}\exp \left\{ \frac{e}{\hbar c}\int_{%
\mathbf{r}_{i}}^{\mathbf{r}_{j}}\mathbf{A}\cdot d\mathbf{l}\right\}
. \label{Hopping}
\end{equation}%
Here $\mathbf{A}=(0,Bx,0)$ is the vector potential of the magnetic
field.

Direct numerical diagonalization of Hamiltonian (\ref{Htb_s}) gives
the field-dependent energy levels $E_{n}(B)$ of electronic states
and corresponding on-site amplitudes of the wave function, $\varphi
_{n,i}$. The orbital magnetic energy of the electronic state at
$T=0$ can be found as function of the chemical potential $\mu $ and
magnetic field $B$ by assumption that all the energy levels below
$\mu $ are double-filled by spin-up and spin-down electrons:

\begin{equation}
U(B,\mu )=2\sum_{n}^{E_{n}<\mu }E_{n}(B),  \label{TotalE}
\end{equation}%
The corresponding orbital susceptibility per unit area and per one
layer is calculated as
\begin{equation}
\chi (\mu )=-\frac{1}{N\sigma }\left[ \frac{\partial ^{2}U(B,\epsilon )}{%
\partial B^{2}}\right] _{B=0},  \label{Sus}
\end{equation}%
where $N$ is the total number of layers and $\sigma
=\sqrt{3}a^{2}n/4$ is the area of a flake containing $n$ carbon
atoms.

\section{Graphene quantum dots}

Before consider  multi-layer clusters we describe the principal
electronic and magnetic properties of single-layer clusters with
zig-zag
edges, studied in\cite%
{Ezawa,Zhang2008,Rossier2007,Wang2008,Wang2009,Potasz,Kosimov,Espinosa,Zarenia,Heiskanen,Espinosa2,Liu,Ominato,Ominato2013}%
.

\begin{figure}[t]
\includegraphics[width=.5\textwidth]{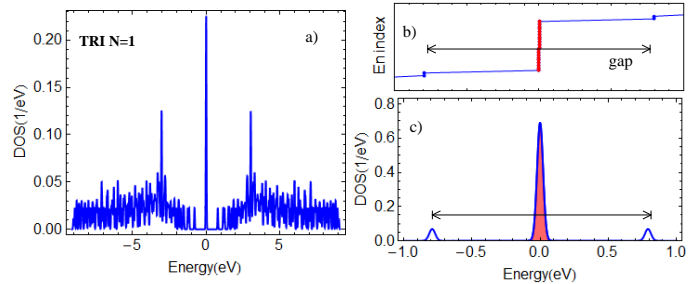}
\caption{Large-energy scale of  DOS of triangular GQD (a), ZES
levels (b) and zoom of DOS(c) within the gap region.} \label{Figt}
\end{figure}

\begin{figure}[h]
\includegraphics[width=.5\textwidth]{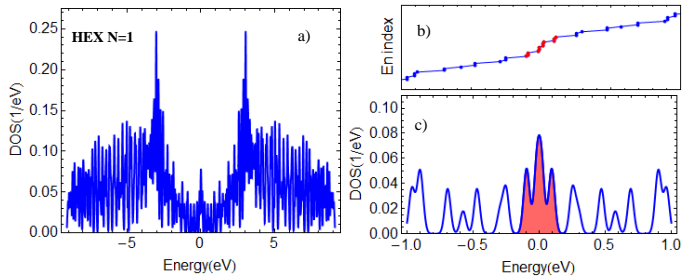}
\caption{DOS of hexagonal GQD (a), the energy levels (b) and the
corresponding DOS in the near-zero energy region, $E\sim 0$ (c). The
levels, corresponded to localized edge-states are noted by red
color.} \label{Figh}
\end{figure}

The DOSs of triangular and hexagonal GQDs with total number of atoms
$n=526$ and $1014$, obtained by diagonalization of TB Hamiltonian
(\ref{Htb_s}) are shown in Fig.\ref{Figt} and Fig.\ref{Figh}. In
general, they repeat the DOS of the infinite graphene layer
\cite{Hobson} smeared by the finite-quantization noise, that
vanishes when size of the cluster increases. The particle-hole
symmetry of DOS,\ $D_{1}(E)=D_{1}(-E)$ is conserved.

\begin{figure}[h]
\includegraphics[width=.5\textwidth]{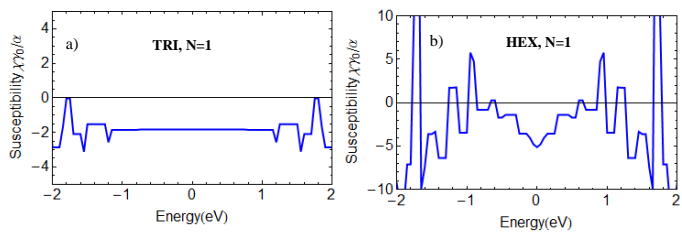}
\caption{Orbital magnetic susceptibility of triangular GQD (a) and
of hexagonal GQD (b).} \label{FigS}
\end{figure}

Fig.\ref{FigS} shows the orbital magnetic susceptibility, obtained
from the magnetic field variation of the energy levels by the
method, described in
Sect. II. The magnetic field was varied between $0\,\mathrm{T}$ and $4\,%
\mathrm{T}$ where the susceptibility was checked to be almost
field-independent. Again, at large energy scale both the
dependencies $\chi (E)$ are similar and are represented by series of
jumps between paramagnetic and diamagnetic values, provided by the
almost-equiprobable up- and downward displacement of the
size-quantized states as function of magnetic field. As cluster size
increases, the magnetic susceptibility for all nanostructures,
disregarding their shape and edge-termination tends to the bulk
limit
characterized by a diamagnetic $\delta $-function singularity at $E=0$. \cite%
{McClure}.

The most important details, distinguishing triangular and hexagonal
GQDs are concentrated at nearly-zero energies when orbital
susceptibility is  diamagnetic.

The DOS of \textit{triangular} GQDs (Fig.\ref{Figt}) reveals the
remarkable feature: the large number of degenerate states is observed \textit{%
exactly} at $E=0$ and is manifested by the huge central peak of zero
energy states (ZES) located inside the energy gap. This property was
shown to be explained by the considerable imbalance of atoms in
cluster sublattices $\emph{A}$ and $\emph{B}$
\cite{Potasz,Rossier2007} that leads to degeneracy

\begin{equation}
\eta _{1}=\sqrt{n+3}-3.  \label{ZESno}
\end{equation}%
The wave functions of ZES are localized mainly at the edges of the
flake \cite{Espinosa2}.

Absence of  electronic states inside the near-zero energy gap
results in
the field-independent diamagnetic plato in $\chi (E)$ at $E\sim 0$ (Fig.\ref%
{FigS}a).\ Importantly, the degenerate states from the central peak
do not contribute to susceptibility since, because of the mentioned
electron-hole symmetry they don't move from their location at $E=0$
when magnetic field is applied.

The level distribution at $E\sim 0$ in \textit{hexagonal} GQDs is
qualitatively different (see Fig.\ref{Figh}). The localized edge
states are not gathered exactly at $E=0$, but are mostly distributed
nearby, inside the band, corresponded to the gap for triangular
clusters. In strike contrast to the triangular case, these
\textit{dispersed }states  give the considerable contribution to the
orbital diamagnetism \cite{Ominato2013,Espinosa2}, demonstrating the
broad diamagnetic peak at $E\sim 0$ in the orbital diamagnetic
susceptibility (Fig.\ref{FigS}b). In general, the diamagnetic
response is larger in hexagonal GQDs as compare to triangular GQDs.

\section{Electronic properties of multilayer quantum dots}

\begin{figure}[t]
\includegraphics[width=.5\textwidth]{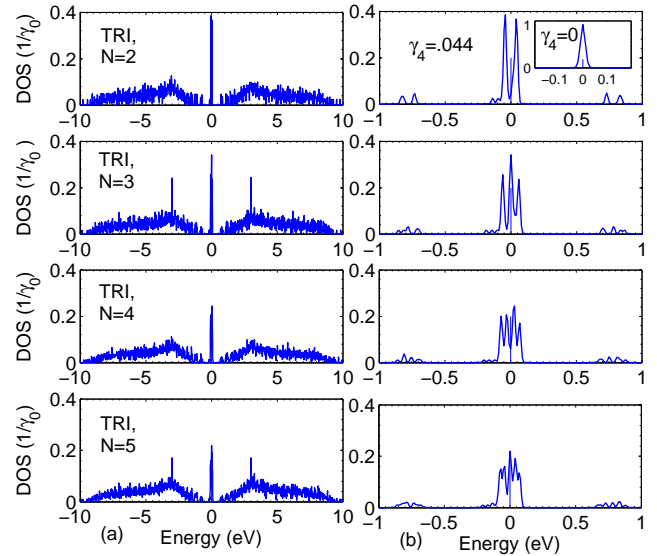}
\caption{DOS of triangular MQD with $N=2\div5$ layers (a) and its
zoom at the near-zero energies (b). The NN coupling parameters are
indicated in the text. The inset shows the unsplit central peak at
$\protect\gamma _{4} =0$.} \label{Fig_DOST}
\end{figure}

We turn now to multilayer clusters with layer number $N=2\div5$. \
The DOSs
of \textit{triangular} MQDs with zig-zag edges are shown in Fig.\ref%
{Fig_DOST}. Similar to the single-layer case the energy gap with the
interior central peak in DOS is observed. The difference however is
that, these states are not located exactly at the Dirac point $E=0$
but smeared around it with formation of $N$ peaks of approximately
the same amplitude (Fig.\ref{Fig_DOST}b). The total number of the
near-zero energy states (NZES) is just the multiple of ZES in each
graphene layer, $\eta _{N}=N\eta _{1}$.

\begin{figure}[h]
\includegraphics[width=.5\textwidth]{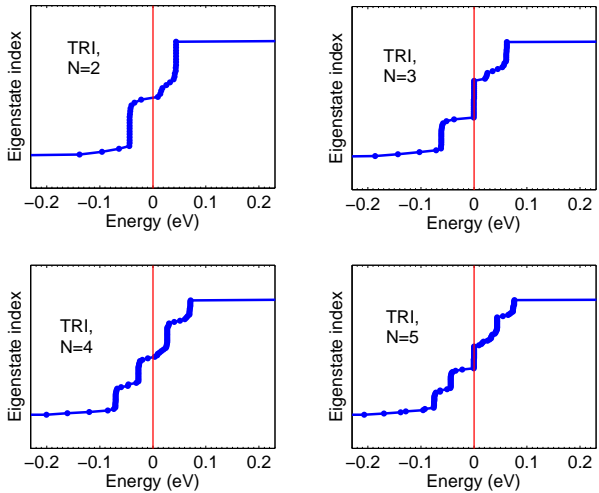}
\caption{ Energy levels for triangular MQD with $N=2\div5$ layers
inside the gap region.} \label{Fig_ET}
\end{figure}

To reveal the detailed structure of NZES we plot the eigenstate
index vs its energy for different $N$ (Fig.\ref{Fig_ET}) and do
observe the energy eigen-level cumulation at the locations of the
split peaks. Importantly, they are not completely degenerate except
the states appeared in the odd-layer clusters exactly at $E=0$. \
Another new property is the
electron-hole asymmetry of DOS with respect to the Dirac point, $%
D_{N}(E)\neq D_{N}(-E)$. To examine the origin of these new features
we tested the variation of DOS under consecutive variation of the
coupling parameters $\gamma _{i}$ and found that this is the
coupling $\gamma _{4}$ which is responsible for the both effects.
Inset to Fig.\ref{Fig_DOST}(b) shows that the central peak is
unsplit at $\gamma _{4}=0$.

Note that $D_{N}(E)$ for MQDs with $N>2$ can be reconstructed from
DOSs of bilayer clusters of the same shape if the dependence of
$D_{2}(E)$ is known
as function of the coupling parameters $\gamma _{1},$ $\gamma _{3}$ and $%
\gamma _{4}$. Generalizing the band decomposition method, proposed
for infinite systems in \cite{Koshino1} we present $D_{N}(E)$ as
\begin{equation}
D_{N}(E)=\frac{1}{2}\sum_{m}D_{2}(E,\lambda _{N,m}\gamma _{1,3,4}),
\label{Decomp}
\end{equation}%
with $m=-(N-1),\,(N-2),\ldots ,N+1$ and with $\gamma$-renormalizing
scaling factors
\begin{equation}
\lambda _{N,m}=2\sin \frac{\left\vert m\right\vert \pi }{2\left(
N+1\right) } \label{mm}
\end{equation}

\begin{figure}[t]
\includegraphics[width=.5\textwidth]{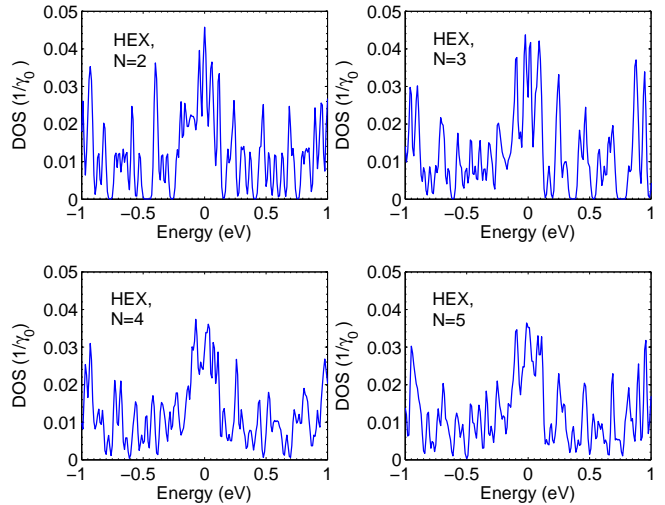}
\caption{ DOS of hexagonal MQD with $N=2\div5$ layers in the
near-zero energy region.} \label{Fig_DOSH}
\end{figure}

Importantly, the term with $m=0$ and $\lambda _{N,0}=0$ exists only
for the odd number of layers. It corresponds to the contribution
from the uncoupled graphene layer, that provides the degenerate ZES
observed in Fig.\ref{Fig_ET}. This residual degeneracy however is
removed when the NNN couplings $\gamma _{2}$ and $\gamma _{5}$ are
taken into account. For even number of layers only two-layers states
contribute to $D_{N}(E)$ and no peaks in DOS appear within the
size-quantization gap that vanishes with increasing of the cluster
size \cite{daCosta}.

\textit{Hexagonal} MQDs, in contrast to triangular MQDs demonstrate
practically the same structure of DOS as the single-layer GQD with
near-zero-energy dispersed electronic states (Fig.\ref{Fig_DOSH}).
The only tiny difference is the electron-hole asymmetry, provided by
the NN\ coupling parameter $\gamma _{4}$.
\begin{figure}[h]
\includegraphics[width=.5\textwidth]{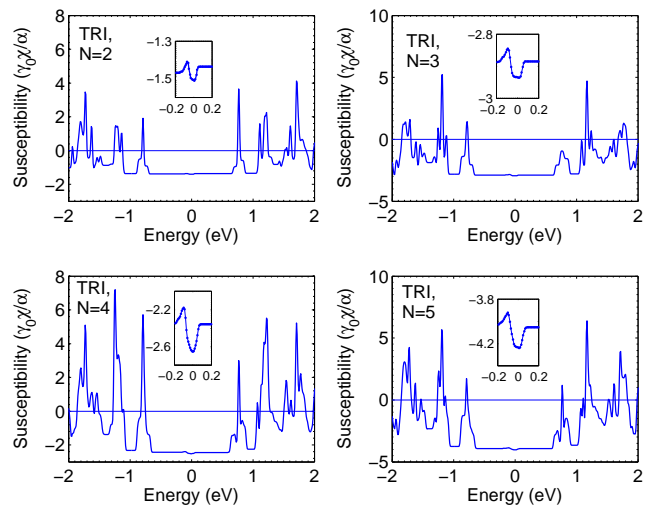}
\caption{Susceptibility of triangular MQD with $N=2\div5$ layers.
Insets show the zoom of asymmetric splash-wavelet feature in the
near-zero energy region.} \label{Fig_ST}
\end{figure}

\section{Magnetic properties of multilayer quantum dots.}

The large-energy scale plot  of magnetic susceptibility  for
\textit{triangular} MQDs with $N=2...5$ demonstrates the
random oscillations between diamagnetic and paramagnetic states (Fig.\ref%
{Fig_ST}), similarly to what was observed for the single-layer case.
Typically, the absolute values of susceptibility for the odd-layer
MQD is always higher than those for the even-layer MQDs that can be
explained by contribution of one decoupled single-layer, whose
susceptibility is remarkably higher \cite{Koshino1}. However at
$E\sim 0$ the prominent asymmetric splash is revealed in the gap
region, where for the $N=1$ case only the flat energy-independent
plato was observed. (Fig.\ref{FigS}a). This feature is provided by
the field-dependent splitting of the central peak due $\gamma _{4}
$-coupling.
\begin{figure}[t]
\includegraphics[width=.5\textwidth]{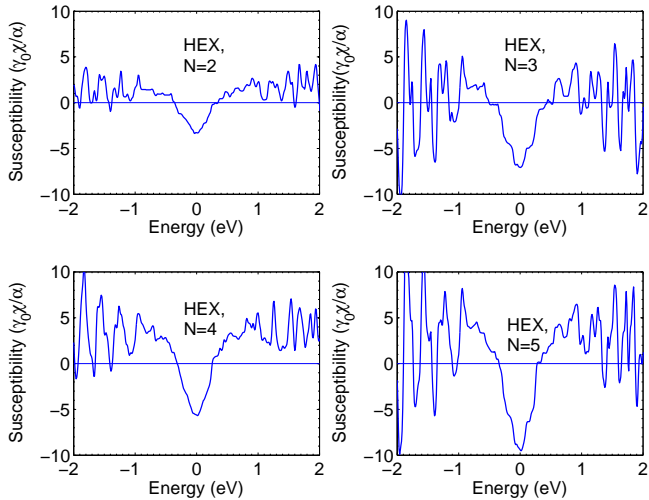}
\caption{Susceptibility of hexagonal MQD with $N=2\div5$ layers.}
\label{Fig_SH}
\end{figure}

The structure of $\chi (E)$ for \textit{hexagonal} MQD is
approximately the same as for hexagonal GQD, albeit some asymmetry
of the broad diamagnetic peak at $E\sim 0$ appears.

\section{Discussion}

In this paper we studied the electronic and magnetic properties of
MQD with zig-zag edges in NN TB approximation as function of the
Fermi energy and their relation with similar properties of GQD. The
behavior of electronic DOS and of orbital magnetic susceptibility in
the near-zero energy region in vicinity of Dirac point is found to
be  provided by the edge-localized electronic states. The details
substantially depend on shape of MQD and on parity of layer number.
In hexagonal MQD the
situation is practically the same as in GQD, previously studied in \cite%
{Espinosa2}: the quasi-continuum distribution of edge-localized
levels is observed at $E\sim 0$ that provides the broad diamagnetic
peak in the orbital susceptibility. In contrast, in triangular MQD
the qualitatively new feature appears. The highly-degenerate ZES
states, centered in the near-zero energy gap-region of triangular
GQD, are split by the inter-layer coupling parameter $\gamma _{4}$
onto the narrow multi-peak band. This gives the non typical
splash-wavelet feature in the orbital diamagnetic susceptibility at
Dirac point, absent for GQD.

In our work we were focused on susceptibility arising from the
orbital electronic properties whereas the spin-paramagnetic effects
were not taken into account. Meanwhile their role can be decisive in
case of highly-degenerate electronic states at $E=0$ in the
half-field triangular GQD with zig-zag edges. The smearing of ZES
into near-zero-energy band in MQD removes such degeneracy and one
can assume that the spin-paramagnetic effects will be less
pronounced there. Meanwhile, this question is less trivial when the
Hubbard-U Coulomb interaction and temperature-induced intra-band
electron jumps are properly taken into account. Therefore study of
the competition between the temperature-independent
orbital-diamagnetic and temperature-dependent spin-paramagnetic
properties in MQD posses the challenging problem for many-body
statistical physics. These effects can be discerned experimentally,
basing on the temperature dependence of susceptibility.

Another interesting property that we observed is the electron-hole
asymmetry with respect to the level with $E=0$, provided by the same
interlayer coupling parameter $\gamma _{4}$. Being of the same
origin as  asymmetric semi-metallic multi-branch  spectrum of
electron-hole carriers in graphite  this feature can result in the
non-zero location of Fermi-level in the half-field MQD. Oscillation
of finite DOS at Fermi-level as function of magnetic field can give
the quasi- de Haas van Alphen oscillations similar to those observed
in the bulk graphite. Study of their character and comparison with
graphite presents another challenging problem.

This work was supported by the Egyptian mission sector and by the
European mobility FP7 Marie Curie programs IRSES-SIMTECH and
ITN-NOTEDEV.

\newpage

\end{document}